\documentclass[prl,twocolumn,letter,longbibliography]{revtex4-1}
\usepackage{amsmath}
\usepackage{amssymb}
\usepackage{graphicx}
\usepackage{hyperref}
\usepackage{braket}
\usepackage{bm}

\def\RuCl{$\alpha$-RuCl$_3$}

\def\ie{{\it i.e.}}

\begin{document}

\title{Confinement transition in a Kitaev-like honeycomb model with
  bond anisotropy}

\author{Gideon Wachtel}
\author{Dror Orgad}
\affiliation{Racah Institute of Physics, The Hebrew University,
  Jerusalem 91904, Israel}

\date{\today}

\begin{abstract}
  The honeycomb $K-\Gamma$ model is known to have both deconfined spin
  liquid and confined phases. We study here a confinement transition
  between the Kitaev spin liquid and a dimerized phase in the limit of
  strong bond anisotropy. By partially projecting out Majorana states
  we are able to map the model onto a model of weakly coupled Ising
  chains in a transverse field. Within this mapping the ordered Ising
  phase corresponds to the condensation of $Z_2$ fluxes, or
  confinement. Our results may improve our understanding of the
  extensively studied spin liquid candidate material \RuCl, where
  $K-\Gamma$ interactions are dominant.
\end{abstract}

\maketitle

One of the defining characteristics of quantum spin liquid phases is
the high level of quantum entanglement in their ground state
\cite{savary_quantum_2017}. These states are commonly associated with
deconfined gauge degrees of freedom on a lattice
\cite{read_large-n_1991, senthil_$z_2$_2000,
  moessner_short-ranged_2001, wen_quantum_2002}. More commonly,
however, spin systems tend to order magnetically, in which cases their
ground states are continuously connected to product states with zero
entanglement entropy. By tuning a given system away from its spin
liquid ground state towards a magnetically ordered state one must
cross a confinement transition followed by breaking of time reversal
symmetry.

A good understanding of such transitions is of great importance in
studying possible spin liquid materials.
One such candidate spin liquid material is \RuCl, a layered honeycomb
Mott insulator which orders magnetically at temperatures well below
its Curie-Weiss scale, most likely due to frustrated magnetic
interactions \cite{sears_magnetic_2015,
  cao_low-temperature_2016}. Furthermore, a broad continuum in
inelastic neutron scattering spectra may indicate the existence of
fractionalized excitations \cite{banerjee_proximate_2016,
  banerjee_neutron_2017, little_antiferromagnetic_2017,
  do_incarnation_2017}. Much of the recent experimental effort has
focused on the intriguing paramagnetic phase obtained by applying an
in-plane magnetic field \cite{sears_phase_2017,
  wolter_field-induced_2017, baek_evidence_2017, zheng_gapless_2017,
  hentrich_unusual_2018, yu_ultralow-temperature_2018,
  kasahara_unusual_2018, shi_field-induced_2018, hentrich_large_2018,
  nagai_two-step_2018}. Due to \RuCl's lattice structure Kitaev
interactions are dominant, and therefore it has been proposed that it
is parametrically close to a Kitaev spin liquid
(KSL) \cite{kitaev_anyons_2006, jackeli_mott_2009}. Specifically, one
scenario suggests that by applying a magnetic field the system
undergoes a deconfinement transition into a spin liquid phase. Kitaev
interactions are not, however, the only dominant spin-spin
interactions, and off-diagonal symmetric, or $\Gamma$
interactions, must be considered at an equal
footing \cite{kim_kitaev_2015, kim_crystal_2016,
  winter_challenges_2016, yadav_kitaev_2016, winter_models_2017,
  lampen-kelley_anisotropic_2018}. Nearest neighbor and longer range
Heisenberg interaction are also present but weaker.

\begin{figure}[t]
  \centering
  \includegraphics[scale=0.4,clip=true]{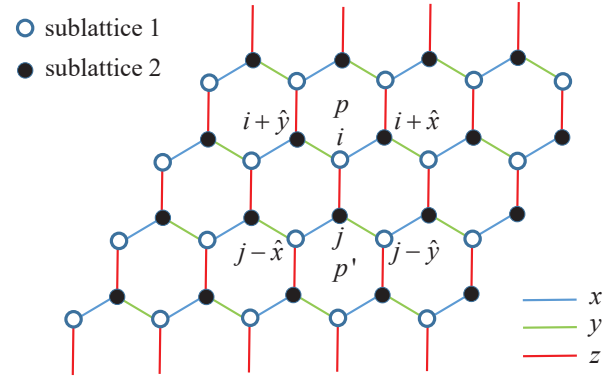}
  \caption{The honeycomb lattice. The $\Gamma_z$ term on $z$-bond
    $\braket{i,j}$ flips one $x$ and one $y$-bond -- either to
    the left or to the right of $\braket{i,j}$. As a result the fluxes
    $w_p$ and $w_{p'}$ are flipped.}
  \label{fig:lattice}
\end{figure}

The Kitaev$\,-\,\Gamma$, or $K-\Gamma$, model is known to exhibit both
a spin liquid phase -- in the pure Kitaev case -- and at least one
magnetically ordered phase \cite{chaloupka_hidden_2015}. The nature of
the two phases is completely different. The spin liquid phase is
characterized by deconfined $Z_2$ gauge degrees of freedom and
fractionalized Majorana excitations \cite{kitaev_anyons_2006}, whereas
the magnetically ordered phase is confined in the gauge theory sense
and its excitations are standard spin waves. Between the two phases
one thus expects a confinement transition, associated with the
condensation of $Z_2$ fluxes. Although the $K-\Gamma$ model has been
studied using a variety of approaches \cite{rau_generic_2014,
  rousochatzakis_classical_2017, winter_breakdown_2017,
  catuneanu_path_2018, gohlke_quantum_2018, winter_probing_2018,
  knolle_dynamics_2018, samarakoon_classical_2018}, there is no known
framework which gives an accurate account of such confinement
transitions. In the following we show that it is possible to work out
the details of a confinement transition in the case of strong bond
anisotropy.  Specifically, we propose to study the anisotropic
$K-\Gamma_z$ model
\begin{equation}
  \label{eq:H}
  H=\sum_{\alpha=x,y,z} K_\alpha \sum_{\braket{i,j}\in\alpha}
  S_i^\alpha S_j^\alpha + \Gamma_z\sum_{\braket{i,j}\in z}
  (S_i^x S_j^y+S_i^y S_j^x)  ,
\end{equation}
where $S_i^\alpha$ are spin-$1/2$ operators located at sites $i$ of a
honeycomb lattice, and $\braket{i,j}$ denote bonds of types $x,y$ or
$z$ extending from a site $i$ on sublattice 1 to site $j$ on sublattice 2,
see Fig. \ref{fig:lattice}. We focus on the case of strong
anisotropy in the Kitaev interaction, $|K_z|\gg~|K_x|,|K_y|$.

In the extreme limit, $K_x = K_y=0$, each of the $z$-bonds is
isolated, with a product state for the ground state. We take $\Gamma_z>0$,
and for the most part consider the case $K_z<0$, where the ground state
is dimerized,
\begin{equation}
  \label{eq:dimer}
  \ket{0}=\prod_{\braket{ij}\in z}\frac{1}{\sqrt{2}}
  \left(\ket{\uparrow\uparrow}_{ij}
  -i\ket{\downarrow\downarrow}_{ij}\right),
\end{equation}
and is obviously confined in the gauge theory sense. However, unlike
the magnetically ordered case, it does not break time
reversal symmetry. This state is stable to small perturbations,
such as small $K_x$ and $K_y$, since it is non-degenerate and gapped
to all excitations.  In the opposite limit, $\Gamma_z=0$, we arrive at
Kitaev's model, which, being in a deconfined phase, has a highly
entangled ground state.  We note that the ground state of the
anisotropic Kitaev model has the same symmetries as the dimer state in
Eq. (\ref{eq:dimer}), and in fact, it is specifically this high
entanglement which differentiates between them.

Our plan is to study the confinement transition between the Kitaev and
dimerized states in the limit of small $K_{x,y}$ and ${\Gamma_z}$
via an effective low-energy Hamiltonian. To this end
we follow Kitaev \cite{kitaev_anyons_2006}, and
rewrite $H$ using four Majorana fermion operators on each site
\begin{equation}
  \label{eq:ibc}
  S_i^\alpha \to \frac{1}{2}ib_i^\alpha c_i.
\end{equation}
Thus,
\begin{eqnarray}
  \label{eq:Ht}
  H\to\tilde H & = & -\frac{1}{4}\sum_{\alpha=x,y,z} K_\alpha\sum_{\braket{i,j}\in\alpha}
  ib_i^\alpha b_j^\alpha\, ic_ic_j \nonumber \\ & &
  -\frac{\Gamma_z}{4}\sum_{\braket{i,j}\in z}
  (ib_i^x b_j^y+ib_i^y b_j^x)\,ic_ic_j.
\end{eqnarray}
The Hamiltonian $\tilde H$ operates in a larger Hilbert space
than that of the original Hamiltonian $H$, a fact reflected by its
gauge symmetry. Explicitly, it is invariant under local
$Z_2$ gauge transformations, $D_i\tilde H D_i= \tilde H$, where
$D_i\equiv b_i^xb_i^yb_i^zc_i$. However, only those states of the
extended Hilbert space $\ket{\psi}$ for which
$D_i\ket{\psi}=\ket{\psi}$ correspond to physical spin states.

The eigenstates of the Kitaev model, defined here by setting $\Gamma_z=0$,
are tensor products $\ket{u}\otimes\ket{\phi_u}$ of states $\ket{u}$
in the $b_i^\alpha$-fermions' sector, and states $\ket{\phi_u}$
in the $c_i$-fermions' sector. The former are simultaneous eigenstates
of the bond operators $ib_i^\alpha b_j^\alpha$, where
$\braket{i,j}\in\alpha$, with eigenvalues $u_{ij}= \pm 1$.
For a given set $u$ of bond values
the $\ket{\phi_u}$ are eigenstates of the free
Majorana Hamiltonian,
\begin{equation}
  \label{eq:Hu}
  H_u=-\frac{1}{4}\sum_\alpha K_\alpha\sum_{\braket{i,j}\in\alpha}u_{ij}ic_ic_j,
\end{equation}
obtained by replacing the $ib_i^\alpha b_j^\alpha$ in $\tilde H$
by $u_{ij}$. In the limit $K_x=K_y=0$ they are eigenstates of the
operators $ic_i c_j$ on $z$-bonds with eigenvalues $\pm1$, and the
degenerate ground-state manifold of $\tilde{H}$ consists of products
$\ket{\psi_u}=\ket{u}\otimes\ket{\phi_u}$
obeying $u_{ij}ic_i c_j=K_z/|K_z|$ for $\braket{i,j}\in z$.
Since excited $c$-fermion states are gapped their effect on the
low-energy physics is captured by standard
degenerate perturbation theory in $K_{x,y}$ and $\Gamma_z$.
The first non-trivial contribution of $K_{x,y}$ to the effective
Hamiltonian reads $H_{\rm eff}(\Gamma_z=0) = -\Delta_v\sum_pw_p$,
where $w_p=\prod_{\braket{i,j}\in p}u_{ij}=\pm 1$ measures the $Z_2$
gauge-invariant flux through a hexagonal plaquette $p$ and
$\Delta_v= K_x^2K_y^2/64|K_z|^3$ is the energy of a vison, \ie, a $w_p=-1$
flux excitation \cite{kitaev_anyons_2006}.

In order to evaluate the leading contribution of the $\Gamma_z$ term to $H_{\rm eff}$
we need its matrix elements between states in the degenerate
manifold. For what follows we find it convenient to fix the gauge and consider
only $\ket{\psi_u}$ for which $u_{ij}=1$ on all $z$-bonds.
%Such an assumption is not restrictive since any
Other states are related to this form by the
action of gauge transformations $D_i$, which change $u_{ij}\to -u_{ij}$
and $c_i\to-c_i$. The invariance $D_i\tilde H D_i= \tilde H$
implies the invariance of the matrix elements under such transformations.
Thus, within the chosen gauge sector, a matrix element of the $\Gamma_z$
term on a given bond $\braket{ij}\in z$ takes the form
\begin{eqnarray}
  \label{eq:G1}
  \Gamma_{ij}(u',u) & \equiv & -\frac{\Gamma_z}{4}
  \bra{\psi_{u'}}(ib_i^x b_j^y+ib_i^yb_j^x)   ic_ic_j\ket{\psi_u}
  \nonumber \\ & = & -\frac{\Gamma_z K_z}{4|K_z|}\braket{u'|ib_i^x b_j^y+ib_i^yb_j^x|u},
\end{eqnarray}
provided that $u$ and $u'$ share the same set of $u_{ij}$ for all $\braket{ij}\in z$,
and vanishes otherwise. Here we have used the condition that determines
the $\ket{\psi_u}$ manifold.

Eq. (\ref{eq:G1}) indicates that while the presence of $\Gamma_z$ interactions does not
change the fact that the $u_{ij}$'s are good quantum numbers for $z$-bonds, it renders the
$u_{ij}$'s on $x$ and $y$-bonds truly dynamical effective spin-1/2 degrees of freedom.
Consequently, it is useful to formulate
$H_{\rm eff}$ in terms of operators acting on the latter. For this purpose we note that
the operator $ib_i^xb_j^y$ operates only on $u_{i,i+\hat{x}}$ and $u_{j-\hat{y},j}$, see
Fig. \ref{fig:lattice}, and that it has the same matrix elements as
%thus allowing us to rewrite it as
$\sigma^2_{i,i+\hat{x}}\sigma^2_{j-\hat{y},j}$,
where $\sigma^2_{kl}$ is the Pauli matrix $\sigma^2$ acting on
$u_{kl}$, and where matrices associated with different bonds commute.
Similarly, $ib_i^yb_j^x$ can be replaced by
$\sigma^2_{i,i+\hat{y}}\sigma^2_{j-\hat{x},j}$.
In addition, Pauli matrices acting on $u_{ij}$
states can also be used to measure the flux on a given plaquette,
namely, $\hat{w}_p=\prod_{\braket{i,j}\in p}\sigma^3_{ij}$. Combining
these results we can cast $H_{\rm eff}$ in the form
\begin{eqnarray}
  \label{eq:Hsigma}
  \!\!\!\!\!\!H_{\rm eff} & =  & -\Delta_v\sum_p\prod_{\braket{i,j}\in p}\sigma^3_{ij}
  % \nonumber \\ & & -\frac{\Gamma_zM}{4}\sum_{\braket{i,j}\in z}
  \nonumber \\ & & -\frac{\Gamma_zK_z}{4|K_z|}\sum_{\braket{i,j}\in z}
  \left(\sigma^2_{i,i+\hat{x}}\sigma^2_{j-\hat{y},j}+
  \sigma^2_{i,i+\hat{y}}\sigma^2_{j-\hat{x},j}\right).%%\nonumber \\
\end{eqnarray}
Further progress is facilitated by employing the constraint
$D_iD_j=1$ for $\braket{i,j}\in z$, which partially projects the model onto
the physical space. In particular, it eliminates states with unphysical
energies on the isolated bond. In terms of the Majorana operators
it becomes
\begin{eqnarray}
  \label{eq:DiDj}
  ib_i^xb_j^y ib_i^yb_j^x  & = & -ib_i^zb_j^zic_ic_j, \nonumber \\
  \sigma^2_{i,i+\hat{x}}\sigma^2_{j-\hat{y},j} \sigma^2_{i,i+\hat{y}}\sigma^2_{j-\hat{x},j}
  & = & -(K_z/|K_z|){\rm I},%%\braket{ic_ic_j},
\end{eqnarray}
where in the last step we have evaluated the operators in the
$\ket{\psi_u}$ subspace.

\begin{figure}[t]
  \centering
  \includegraphics[scale=0.4,clip=true]{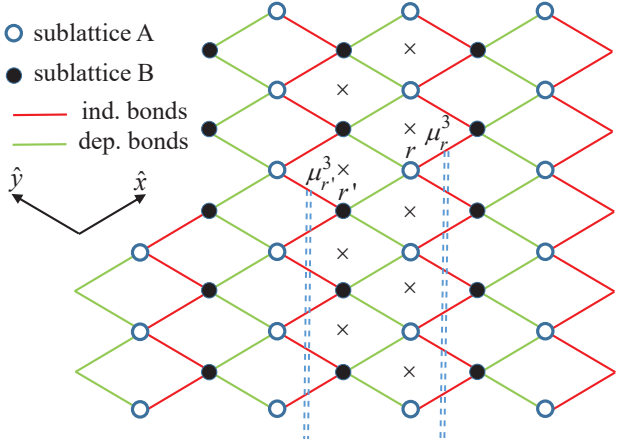}
  \caption{The rhombic lattice obtained by collapsing the $z$-bonds
    onto points. Independent bond degrees of freedom are denoted in red,
    while dependent ones are shown in green. The double-dashed lines correspond
    to strings of independent $\sigma^2$ operators which enter $\mu_r^3$ and
    $\mu_{r'}^3$, where $r\in A$ and $r'\in B$ sublattices. Few sites of
    the dual lattice are indicated by $\times$'s.}
  \label{fig:duallat}
\end{figure}
At this point it is possible to collapse the $z$-bonds onto
points on a rhombic lattice, and consider the $u_{ij}$ degrees of
freedom on the remaining $x$ and $y$-bonds, see
Fig. \ref{fig:duallat}. Denoting the points on the new lattice by $r$
we recast $H_{\rm eff}$ into
\begin{eqnarray}
  \label{eq:Hr}
  H_{\rm eff} & = & -\Delta_v\sum_r\sigma^3_{r,r+\hat{x}}
  \sigma^3_{r+\hat{x},r+\hat{x}+\hat{y}}
  \sigma^3_{r+\hat{y},r+\hat{y}+\hat{x}}\sigma^3_{r,r+\hat{y}}
  % \nonumber \\ & & -\frac{\Gamma_zM}{4}%%\braket{ic_ic_j}
  \nonumber \\ & & -\frac{\Gamma_zK_z}{4|K_z|}%%\braket{ic_ic_j}
  \sum_r\left(\sigma^2_{r,r+\hat{x}}\sigma^2_{r,r-\hat{y}}+
    \sigma^2_{r,r+\hat{y}}\sigma^2_{r,r-\hat{x}}\right),
  %%\nonumber \\
\end{eqnarray}
where each site $r$ has four connected bonds, denoted by
$r,r\pm\hat{x}$ and $r,r\pm\hat{y}$.  The constraint,
Eq. (\ref{eq:DiDj}), now becomes
\begin{equation}
  \label{eq:Gr}
  \sigma^2_{r,r+\hat{x}}\sigma^2_{r,r-\hat{y}}
  \sigma^2_{r,r+\hat{y}}\sigma^2_{r,r-\hat{x}}
  = -(K_z/|K_z|){\rm I}.%%\braket{ic_ic_j}.
\end{equation}
Within this constraint, half of the bonds can be treated as
dependent degrees of freedom \cite{kogut_introduction_1979}. Consider,
for example, site $r$ in Fig. \ref{fig:duallat}. We can use
Eq. (\ref{eq:Gr}) to express $\sigma^2_{r,r+\hat{y}}$ in terms
of $\sigma^2_{r,r+\hat{x}}$, $\sigma^2_{r,r-\hat{y}}$ and
$\sigma^2_{r,r-\hat{x}}$. In turn, the constraint on site $r'=r-\hat{x}$
implies that $\sigma^2_{r,r-\hat{x}}$ is given by $\sigma^2_{r,r-\hat{x}+\hat{y}}$,
$\sigma^2_{r,r-2\hat{x}}$ and $\sigma^2_{r,r-\hat{x}-\hat{y}}$,
and so forth. Consequently, for all points $r$ on sublattice $A$,
see Fig. \ref{fig:duallat}, $\sigma^2_{r,r+\hat{y}}$ and $\sigma^2_{r,r-\hat{x}}$
are expressible as products of $\sigma^2_{r,r+\hat{x}}$ and $\sigma^2_{r,r-\hat{y}}$
on other sites of the $A$ sublattice. Hence, we take the former to be dependent on
the latter. The roles are reversed for points $r$ on sublattice $B$.
Once the dependent degrees of freedom have been eliminated in favor of
the independent bonds, the effective Hamiltonian commutes with $\sigma^3$
on the dependent bonds and we can simply set them to $\sigma^3=1$.
The result for $K_z<0$ is
\begin{eqnarray}
  \label{eq:HGr}
  H_{\rm eff} & = & -\sum_{r\in A}\left[\Delta_v\sigma^3_{r,r+\hat{x}}
    \sigma^3_{r+\hat{x},r+\hat{x}+\hat{y}}-\frac{\Gamma_z}{2}\sigma^2_{r,r+\hat{x}}
    \sigma^2_{r,r-\hat{y}}\right] \nonumber \\
  & & -\sum_{r\in B}\left[\Delta_v
    \sigma^3_{r,r+\hat{y}}\sigma^3_{r+\hat{y},r+\hat{y}+\hat{x}} -\frac{\Gamma_z}{2}\sigma^2_{r,r+\hat{y}}
    \sigma^2_{r,r-\hat{x}}\right]. \nonumber \\
\end{eqnarray}
When $K_z>0$, the $\Gamma_z$ term is expected to additionally scale as
$(K_{x,y}/K_z)^4$ since lower order terms cancel each other when
applying Eq. (\ref{eq:Gr}) to Eq. (\ref{eq:Hr}).

The final stage of the calculation consists of a duality
transformation which maps $H_{\rm eff}$ onto a model of disconnected
quantum Ising chains, with spin-1/2 degrees of freedom on the points
of the dual lattice, see Fig. \ref{fig:duallat}. To do so we define
a new set of dual spin-flip operators in terms of
the independent bond operators
\begin{equation}
  \label{eq:mu2}
  \mu^1_r = \left\{
    \begin{array}{ccc}
      \sigma^3_{r,r+\hat{x}}\sigma^3_{r+\hat{x},r+\hat{x}+\hat{y}} & & r\in A \\ \\
      \sigma^3_{r,r+\hat{y}}\sigma^3_{r+\hat{y},r+\hat{y}+\hat{x}} & & r\in B
    \end{array}\right. .
\end{equation}
These two sets of operators correspond to the $e$ and $m$ excitations
in Kitaev's work\cite{kitaev_anyons_2006}. At the same time, $\mu^3_r$
is defined as a product of $\sigma^2$ operators on a semi-infinite
string, running parallel to the original $z$-bonds, and which
terminates at the plaquette above $r$
\begin{equation}
  \label{eq:mu3}
  \mu^3_r = \left\{
    \begin{array}{ccc}
      \prod_{l\ge 0}\sigma^2_{r-l\hat{a},r+\hat{x}-l\hat{a}}
      \sigma^2_{r-l\hat{a},r-\hat{y}-l\hat{a}} & & r \in A  \\ \\
      \prod_{l\ge 0}\sigma^2_{r-l\hat{a},r+\hat{y}-l\hat{a}}
      \sigma^2_{r-l\hat{a},r-\hat{x}-l\hat{a}} & & r \in B
    \end{array}
  \right. ,
\end{equation}
see Fig. \ref{fig:duallat}. Here we have defined the displacement
parallel to the strings as $\hat{a}=\hat{x}+\hat{y}$. It is
possible to verify that the dual operators obey the same
(anti-)commutation relations as spin-1/2 operators,
$\{\mu_r^\alpha,\mu_{r}^\beta\} = 2\rm I\delta_{\alpha\beta}$, and
$[\mu_r^\alpha,\mu_{r'}^\beta] = 0$ for $r\ne r'$.  Noting that
\begin{equation}
  \label{eq:mumu}
  \mu^3_r\mu^3_{r-\hat{a}} = \left\{
    \begin{array}{ccc}
      \sigma^2_{r,r+\hat{x}}\sigma^2_{r,r-\hat{y}} & & r \in A  \\ \\
      \sigma^2_{r,r+\hat{y}}\sigma^2_{r,r-\hat{x}} & & r \in B
    \end{array}
  \right. ,
\end{equation}
we finally obtain the dual Hamiltonian
\begin{equation}
  \label{eq:Hd}
  H_d = \frac{\Gamma_z}{2}\sum_r\mu^3_r\mu^3_{r-\hat{a}}
  -\Delta_v\sum_r\mu^1_r,
\end{equation}
describing a set of decoupled quantum Ising chains. When $\Gamma_z/2 >
\Delta_v$ this dual model is known to transition into an ordered
phase \cite{pfeuty_one-dimensional_1970}, which corresponds to
condensation of visons in the original model, and therefore, to
confinement. For $K_z<0$ this happens at
\begin{equation}
  \label{eq:Gc}
  \frac{\Gamma_{z,c}}{|K_z|}\approx\frac{2\Delta_v}{|K_z|}\approx
      \frac{1}{32}\left(\frac{K_{x,y}}{K_z}\right)^4.
\end{equation}

\begin{figure}[t]
  \centering
  \includegraphics[width=\linewidth]{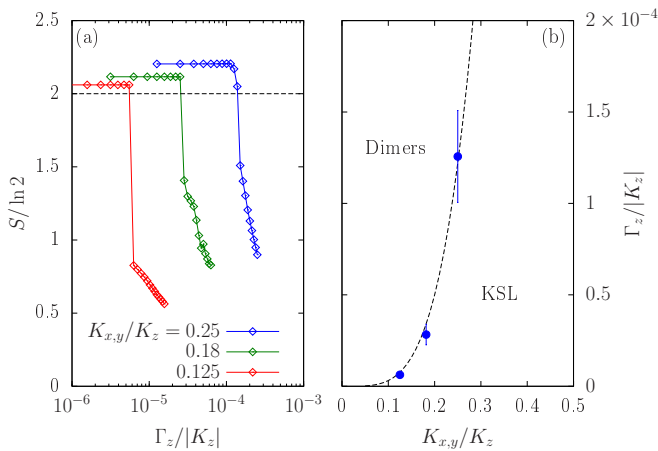}
  \caption{(a) Entanglement entropy calculated using DMRG for a
    bipartition cutting only $x$ and $y$-bonds. In the deconfined
    phase $S$ is slightly above the dashed $2\ln 2$ line. (b)
    Confinement transition curve based on Eq. (\ref{eq:Gc}), with
    points obtained from (a).}
  \label{fig:DMRG}
\end{figure}
This is the main result of our calculation. To support it we performed
density-matrix renormalization-group (DMRG) \cite{white_density_1992}
calculations of the $K-\Gamma_z$ model on a cylindrical $6\times16$ site
system and calculated the entanglement entropy $S$ for a bi-partition cutting
only $x$ and $y$-bonds. In the dimerized phase, for weak $K_{x,y}$,
the $z$-bond dimers are only weakly entangled and $S$ is expected to be small.
In the deconfined phase, however, there is a contribution to the
entanglement entropy coming from the gauge sector. Specifically, for
the Kitaev honeycomb spin liquid the gauge sector ($b$-fermions)
entanglement entropy for a bi-partition cutting $2L$ bonds is given by
\cite{yao_entanglement_2010, dora_gauge_2018} $S_G\simeq (L-1)\ln
2$. Fig. \ref{fig:DMRG}a shows $S$ as a function of $\Gamma_z$ for
different values of $K_{x,y}$. At low $\Gamma_z$ $S$ is slightly above
$2\ln 2$ as expected for a Kitaev spin liquid on a cylinder with a 6
bond circumference. The excess entropy comes from the gapped
$c$-fermions. A confinement transition to the dimerized state is seen
as a drop in $S$, and occurs at lower $\Gamma_z$ as $K_{x,y}$ is
decreased. The critical values of $\Gamma_z$ are plotted in
fig. \ref{fig:DMRG}b, and approximately follow Eq. (\ref{eq:Gc}), in
support of our calculation. In the following we discuss some
generalizations, which go beyond our model, Eq. (\ref{eq:H}), in its
strong anisotropy limit.

\emph{Tuning towards the isotropic limit.} As $K_{x,y}$ increase,
terms of orders higher than $(K_{x,y}/K_z)^4$ must be included in
Eq. (\ref{eq:Hsigma}). They translate into $\mu_r^1\mu_{r'}^1$ terms,
which decrease with $|r-r'|$ and introduce weak interactions between
the otherwise decoupled Ising chains in Eq. (\ref{eq:Hd}). Thus, the
critical behavior is expected to belong to the three-dimensional Ising
universality class.
Since a pair of neighboring visons have a lower energy than the two
isolated visons, it is likely that as the model is tuned towards the
isotropic point the critical $\Gamma_z$ will be lower than twice the
bare vison gap $2\Delta_v$. Note, however, that $\Delta_v$ itself
grows rapidly with $K_{x,y}$, making the deconfined region larger in
the isotropic limit.
Eventually, the $c$-fermions become gapless and our treatment breaks down.

\emph{Other spin-spin interactions.} The Heisenberg interaction
$J(S_i^x S_j^x+S_i^yS_j^y)$ on $z$-bonds can be treated on equal
footing to the $\Gamma_z$ term. Its order $J$ contribution to $H_{\rm eff}$
vanishes within the physical space when $K_z<0$, but survives
for $K_z>0$ and couples the Ising chains. On the other
hand, $\Gamma$ and Heisenberg terms on $x$ and $y$-bonds do not
preserve the choice of gauge, which we have used, and therefore
lie outside the scope of the present framework.

\emph{Effect of a magnetic field.} Consider, for example, the
perturbation $-h\sum_i(S_i^x+S_i^y)$ due to a magnetic field
$\vec{h}=(h,h,0)$.  For $K_z<0$, and to leading order in $h$ within
the physical space it shifts the strength of the $\Gamma_z$ term to
$\Gamma_z-4h^2/|K_z|$. Thus, when starting in the $\Gamma_z>2\Delta_v$
confined dimerized phase, increasing the magnetic field suppresses the
$\Gamma$ term and may induce a deconfinement transition.  We note that
this effect, of order $h^2$, is more relevant than the time reversal
symmetry breaking $h^3$ terms considered in studies focusing on a
possible chiral spin liquid phase in
\RuCl~\cite{gohlke_dynamical_2018, ronquillo_signatures_2018,
  zhu_robust_2018, mcclarty_topological_2018, nasu_successive_2018,
  jiang_field_2018, zou_field-induced_2018}.

\emph{Conclusion.} Starting from a model of anisotropic local
spin interactions on a honeycomb lattice, we determined the phase
boundary between a deconfined KSL phase and a confined dimerized
phase. It is difficult to directly apply our results to \RuCl~since
the bond anisotropy is not expected to be as large as considered
here. Furthermore, non-negligible Heisenberg interactions probably
exist in the material, stabilizing a magnetically-ordered confined
phase rather than the dimerized confined phase of the model above. A
dimerized state has been observed in \RuCl~under pressure
\cite{bastien_pressure-induced_2018, li_raman_2018,
  yadav_strain-_2018}, but it is most likely not the same as in
Eq. (\ref{eq:dimer}). Nevertheless, we do expect the framework outlined
here to be a good starting point for more detailed discussions of
confinement transitions in \RuCl~and similar materials.

We would like to thank Y. B. Kim for discussions which led to this
work.

\bibliography{kgammamod}

\end{document}